\documentclass[%
 reprint,
 amsmath,amssymb,
 aps,
prb,
floatfix
]{revtex4-1}

\usepackage{graphicx}
\usepackage{dcolumn}
\usepackage{bm}
\usepackage[english]{babel}
\usepackage{color}
\usepackage{amssymb,amsmath,color,times}
\usepackage[colorlinks=true,linkcolor=blue,citecolor=blue]{hyperref}

\begin{document}

\preprint{APS/123-QED}

\title{Observation of extremely long exciton lifetime in Janus-MoSTe monolayer}
\author{Hao Jin}%
\author{Tao Wang}
\author{Zhi-Rui Gong}%
\email{gongzr@szu.edu.cn}
\affiliation{
College of Physics and Energy, Shenzhen Key Laboratory of Advanced Thin Films and Applications, Shenzhen University, Shenzhen 518060, China.
}%

\author{Chen Long}
\author{Ying Dai}%
\affiliation{%
 School of Physics, State Key Laboratory of Crystal Materials, Shandong University, Jinan 250100, China.
}%

\date{\today}

\begin{abstract}
The electron-hole separation efficiency is a key factor that determines the performance of two-dimensional (2D) transition metal dichalcogenides (TMDs) and devices. Therefore, searching for novel 2D TMD materials with long timescale of carrier lifetime becomes one of the most important topics. Here, based on the time-domain density functional theory (TD-DFT), we propose a brand new TMD material, namely janus-MoSTe, which exhibits a strong build-in electric field. Our results show that in janus-MoSTe monolayer, the exciton consisting of electron and hole has a relatively wide spatial extension and low binding energy. In addition, a slow electron-hole recombination process is observed, with timescale on the order of 1.31 ns, which is even comparable with those of van der Waals (vdW) heterostructures. Further analysis reveals that the extremely long timescale for electron-hole recombination could be ascribed to the strong Coulomb screening effect as well as the small overlap of wavefunctions between electrons and holes. Our findings establish the build-in electric field as an effective factor to control the electron-hole recombination dynamics in TMD monolayers and facilitate their future applications in light detecting and harvesting.

\end{abstract}

\maketitle


\section{Introduction}
Since the successful exfoliation of graphene, there has been growing interest in atomically two-dimensional (2D) materials.\cite{RN724,hernandez2008high}  Among them, particular attention has been paid to 2D transition metal dichalcogenides (TMDs) due to their promising properties, which hold potential applications in electronic and optoelectronic devices.\cite{Mark10,MoS212,xia2014rediscovering} However, due to the large effective masses and weak dielectric screening, 2D TMDs display strong excitonic effect with high exciton binding energies.\cite{Ye2014Probing,PENG2015128} As a result, the recombination rate of photongenerated electron and hole is usually very fast, in the timescale of several hundreds of picoseconds.\cite{Korn2011,Sun2014,Linqiu17} The short lifetime for electron-hole pair would dramatically impede the performance of 2D TMD materials and devices, leading to extremely low quantum efficiency, i.e. in the 10$^{-4}$ $\sim$ 10$^{-2}$ range.\cite{sundaram2013electroluminescence,ross2014electrically,Wang2015Ultrafast}

Recently, a new kind of layered TMD material namely janus-MoSSe has been successfully synthesized by chemical vapor deposition (CVD) method, in which the top layer of Se atoms in MoSe$_2$ monolayer is replaced by S atoms, while the bottom Se layer remaining intact.\cite{lu2017janus,zhang2017janus} In comparison with regular MoSe$_2$, the janus-MoSSe monolayer is reported to have out-of-plane build-in electric field, which opens up a way to tune the properties of TMD monolayers.\cite{li2017electronic} It should be pointed out that due to the strong excitonic effect, the recombination rate of electron-hole pair in janus-MoSSe monolayer is still too fast to meet the desirable applications. Thus, searching for new 2D TMD materials with long timescale for electron-hole recombination becomes a crucial scientific issue.

In this work, based on the time-domain density functional theory (TD-DFT), we propose a brand-new TMD material, namely janus-MoSTe. We find that due to the discrepancy of electronegativity between S and Te atoms, janus-MoSTe monolayer exhibits large build-in electric field, which significantly weakens the binding strength of excitons. The calculated exciton binding energy is as low as 0.54 eV according to GW/Bethe-Salpeter Equation (BSE). The electron-hole recombination dynamics is then explored in details based on the nonadiabatic molecular dynamics (NAMD) technique. A slow electron-hole recombination process is observed in monolayer janus-MoSTe, with the lifetime up to 1.31 ns, which is even comparable with those of van der Waals (vdW) heterostructures made by stacking layered TMDs.\cite{Rivera2015Observation,Li2016Charge} Such extremely long exciton lifetime is interpreted by the strong Coulomb screening effect as well as the the greatly reduced overlap of the out-of-plane wavefunctions between the electron and hole, which both result from the large build-in electric field. These results suggest that the proposed janus-MoSTe monolayer has great potentials in further optoelectronic and photovoltaic applications.

\section{Methodologies}
The DFT calculations are carried out using the projector augmented wave (PAW)\cite{kresse1999ultrasoft} method and plane-wave basis set as implemented in the Vienna $ab\ initio$ simulation package (VASP).\cite{KressePRB96, Kresse96} The Perdew-Burke-Ernerhof (PBE) parametrization of the generalized gradient approximation (GGA) is used for the exchange-correlation potentials with a cutoff energy of 500 eV.\cite{perdew1996generalized} A $18 \times 18 \times 1$ \emph{k}-point mesh is employed to sample the first Brillouin zone. The vdW correction to the GGA functional based on Grimme's scheme is incorporated to better describe the nonbonding interaction.\cite{Grimme06}  All atoms are fully relaxed until a precision of 10$^{-5}$ eV in energy and 0.01 eV/\r{A} in residual forces. The state-of-the-art hybrid functional (HSE06) is used, in which the hybrid functional is mixed with 25\% exact Hartree-Fock (HF) exchange.\cite{Heyd03,Paier06} In order to investigate the optical properties of TMDs heterostructures, a many-body GW+BSE calculation is performed.\cite{PhysRevB.62.4927,PhysRevB.74.035101} The number of empty bands and the cutoff energy for response function are tested to ensure that the gaps are converged with an accuracy of 0.01 eV. A vacuum space of at least 18 \AA\ is added in the vertical direction to ensure decoupling between adjacent layers.\cite{olsen2011dispersive}
\begin{figure*}
\centering
\includegraphics[width=6in]{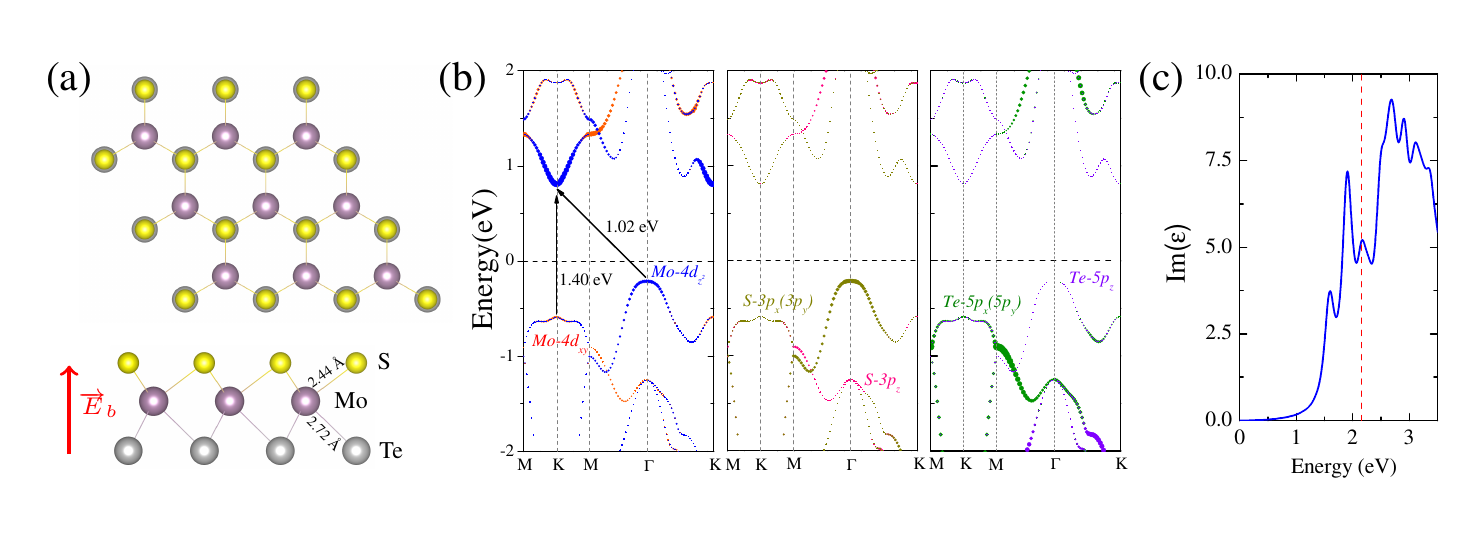}
\caption{(a) Top and side views of the janus-MoSTe supercell. (b) Orbit-projected band structures of janus-MoSTe monolayer at the PBE level. The Fermi level is set to be zero. (c) G$_0$W$_0$+BSE adsorption spectrum for janus-MoSTe monolayer. The vertical dash line represents bandgap at the G$_0$W$_0$ level (see Supplementary Figure 1b).}
\label{fig:ele}
\end{figure*}

The photoinduced current is carried out in the framework of DFT method combined with the nonequilibrium Green's functions (NEGF-DFT).\cite{nanodcal1,nanodcal2} The photocurrent is determined as:\cite{HaugQuantum,Chen2012First,Zhang2014Generation}
\begin{equation}
J_{ph}=\frac{e}{2\hbar}\int \frac{dE}{2\pi} \sum_\alpha T_{\alpha}(E)
\end{equation}
where $\alpha$ stands for the lead of source/drain. $T_{\alpha}(E)$ is the effective transmission coefficient of lead $\alpha$.\cite{HaugQuantum,JinJMCC}

The NAMD simulations are carried out using PYXAID software package,\cite{Akimov2013The,Akimov2014Advanced} which uses decoherence induced surface hopping (DISH) technique to deal with the electron-hole recombination process.\cite{Jaeger2012Decoherence} The time-dependent Kohn-Sham equations are solved along the atomic trajectory $\mathbf{R}$, which is obtained from the $ab$ $initio$ adiabatic molecular dynamics (AMD) results. A detailed description of the algorithm can be found in previous work.\cite{Akimov2013The,Linqiu17,Long2017Nonadiabatic} The janus-TMDs monolayers are heated to 300 K, and 5 ps AMD trajectories with a 1 fs atomic time-step are produced, which are employed to calculate the nonadiabatic Hamiltonians (NAHs). The 5 ps NAHs are then iterated 20 times to simulate the recombination dynamics over a long period. 200 initial conditions from the first 5 ps of NAHs are chosen to sample the canonical distribution of the atomic coordinates. NAMD simulations are then performed starting from each initial condition and using 4000 random number sequences to compute the surface hopping probabilities.

\section{Results and Discussion}
As shown in \autoref{fig:ele}(a), the so-called janus-MoSTe has a sandwiched structure, namely S-Mo-Te from the top to the bottom. The lattice parameter is calculated to be 3.37 \AA, and the bond lengths of Mo-S and Mo-Te are 2.44 and 2.72 \AA, respectively. The asymmetric geometry results in a dipole moment up to 0.38 Debye along the vertical direction, which is about 60\% larger than that in janus-MoSSe monolayer. Based on the orbital characteristics, the band structures are plotted using different colors. As shown in \autoref{fig:ele}(b), monolayer janus-MoSTe is an indirect bandgap semiconductor, with the bandgap of 1.02 eV at the PBE level. The conduction band minimum (CBM) occurs at $K$ point, which is dominated by Mo-4$d_{z^{2}}$ orbital. While the valence band maximum (VBM) locates at $\Gamma$ point, which is composed of Mo-4$d_{xy}$ orbital as well as S-3$p_{x}$ (3$p_{y}$) and Te-5$p_{x}$ (5$p_{y}$) orbitals. Note that the calculations at the PBE level often underestimate the bandgaps of the semiconductors. To obtain more accurate band structures, HSE06 method is employed in this work. As shown in Supplementary Figure 1a, the bandgap at the HSE06 level is obviously larger than that from PBE, with the value up to 1.84 eV. Nevertheless, the dispersion of the band structure exhibits similar tendency in both VBM and CBM.
\begin{table}[t]
\caption{The calculated dipole moment ($p$), static dielectric constant ($\varepsilon$), exciton Bohr radius ($a^*$), exciton binding energy ($E_b$), and electron-hole recombination lifetime ($\tau$) for three TMDs monolayers.}
\begin{tabular}{lccccc}
\hline
\hline
             & $p$ (Debye)   & $\varepsilon$& $a^{*}$(\AA) & $E_{b}$ (eV) & $\tau$ (ns) \\

\hline
    MoS$_2$  & --   & 8.05  & 10.3 & 0.80 & 0.39\cite{Linqiu17}\\
    janus-MoSSe    & 0.24 & 8.67  & 13.2 & 0.63 & 0.44\\
    janus-MoSTe    & 0.38 & 9.63  & 17.1 & 0.54 & 1.31\\
\hline
\end{tabular}
\label{tab:table1}

\end{table}

In the following, we look into the nature of exciton in janus-TMDs monolayer influenced by the build-in electric filed. Using hydrogenic model, we first estimate the exciton Bohr radius ($a^{*}$), which is expressed as:\cite{dvorak2013origin}
\begin{equation}
a^{*} = \frac{m_{0}}{\mu}\varepsilon a_{H}
\end{equation}
where $m_{0}$ , $\mu$ and $a_{H}$ are the free electron mass, effective reduced mass, and Bohr radius of the hydrogen atom, respectively. $\varepsilon$ denotes the macroscopic dielectric constant. The calculated static dielectric constant ($\varepsilon$) and exciton Bohr radius ($a^*$) are given in \autoref{tab:table1}. Clearly, the large value of dipole moment in MoSTe leads to the strong Coulomb screening effect, which gives rise to a relatively broad spatial extension namely larger Bohr radius.

With respect to the optical absorption properties of the monolayers, $e-h$ interactions can be taken into account with the BSE approach based on the quasiparticle corrections. In \autoref{fig:ele}c, we plot the imaginary part of the transverse dielectric constant. The exciton binding energy is then calculated, which is defined as the difference between the GW bandgap and the first absorption peak of the spectrum. The calculated exciton binding energies of janus-MoSTe is as low as 0.54 eV, which is significantly reduced as compared with MoS$_2$ monolayer (0.80 eV). This can be understood based on the fact that the build-in electric field enhances screening potential, leading to low excitonic binding energy.

\begin{figure}[t]
  \centering
\includegraphics[width=3in]{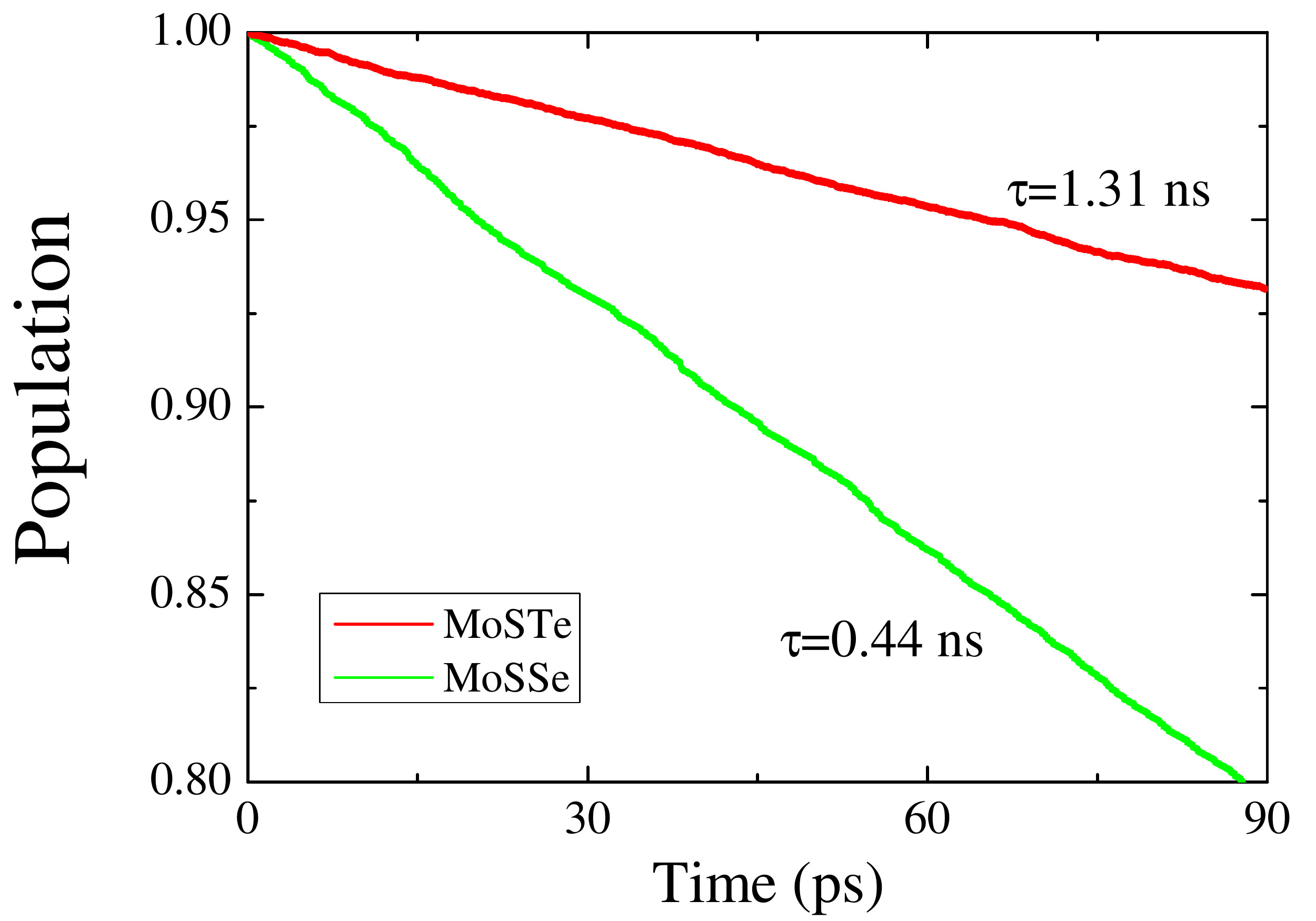}
\caption{Electron-hole recombination dynamics for janus-MoSSe and janus-MoSTe monolayers.}
\label{fig:pop}
\end{figure}

To explore the influence of the build-in electric filed on the charge transfer dynamics, we investigate the electron-hole recombination kinetics based on the TD-DFT. To gain quantitative information on recombination dynamics, we fit the time evolution data with an exponential equation $f(t) = A_0 + B_0exp (-t/\tau)$, where $\tau$ is the recombination lifetime. The results are characterized in \autoref{fig:pop}. It is clearly that the obtained lifetime in janus-MoSSe monolayer is 0.44 ns, which is in the same time scale as compared with the value in MoS$_2$ monolayer (see \autoref{tab:table1}).\cite{Linqiu17} As a result, the recombination rate in janus-MoSSe monolayer is still too fast for further applications. By contrast, the obtained time scale in janus-MoSTe is extremely long. The introduction of build-in electric field in the janus-MoSTe monolayer dramatically reduces the recombination rate, resulting in a longer excited-state lifetime of 1.31 ns, which is even comparable with the time scales in those TMDs heterostructures, i.e. 1.4 $\sim$ 1.8 ns. \cite{Rivera2015Observation,Li2016Charge}

\begin{figure}[t]
  \centering
\includegraphics[width=3in]{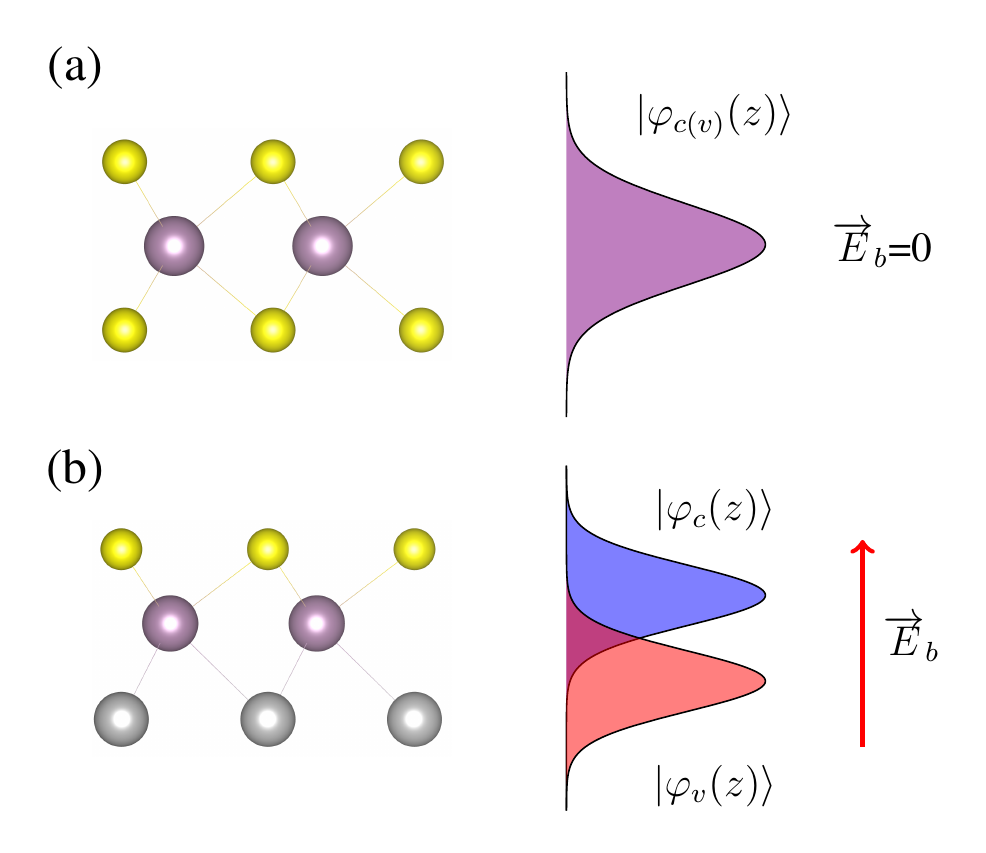}
\caption{Schematic diagram of the out-of-plane wavefunction for (a) $\mathrm{MoS_{2}}$ and (b) $\mathrm{MoSTe}$. The out-of-plane wavefunctions of electrons and holes are respectively represented by envelopes filling by the blue and red colors. The overlap of them are represented by the envelope filling by the purple color. The strong build-in electric field in $\mathrm{MoSTe}$ seperates the center of the out-of-plane wavefunctions $\varphi_{c\left(v\right)}\left(z\right)$ and significantly reduces the overlap of out-of-plane wavefunction. This mechanism vanishes in $\mathrm{MoS_{2}}$ because there is no build-in electric field. }
\label{fig:demo}
\end{figure}

It is worthwhile to study the mechanism by which the electron-hole pair excitation in janus-TMDs monolayers. The lifetime of the exciton consisting of electron and hole is inversely
proportional to the overlap of their wavefunction as $\tau\propto\left|\left\langle \phi_{c}\left(\mathbf{r}\right)|\phi_{v}\left(\mathbf{r}\right)\right\rangle \right|^{-2}$, where c and v respectively denote the conduction and valence band. Such overlap actually represents the probability of the electron and
hole to meet each other and recombine, which can be significantly
reduced by the out-of-plane build in electric field $\overrightarrow{E_{b}}$.
For the sake of simplicity, we assume the electron (hole) wavefunction
is a product state of the in-plane part $\psi_{c\left(v\right)}\left(x,y\right)$
and the out-of-plane part $\varphi_{c\left(v\right)}\left(z\right)$
as $\phi_{c\left(v\right)}\left(\mathbf{r}\right)=\psi_{c\left(v\right)}\left(x,y\right)\varphi_{c\left(v\right)}\left(z\right)$.
In this sense, the overlap of wavefunction becomes
\begin{eqnarray}
 &  & \left|\left\langle \phi_{c}\left(\mathbf{r}\right)|\phi_{v}\left(\mathbf{r}\right)\right\rangle \right|^{-2} \nonumber \\
 & = & \left|\left\langle \psi_{c}\left(x,y\right)|\psi_{v}\left(x,y\right)\right\rangle \right|^{-2}\left|\left\langle \varphi_{c}\left(z\right)|\phi_{v}\left(z\right)\right\rangle \right|^{-2} \nonumber \\
 & = & \frac{\pi}{2}\left(a^{*}\right)^{2}\left|\left\langle \varphi_{c}\left(z\right)|\phi_{v}\left(z\right)\right\rangle \right|^{-2},
\label{equ1}
\end{eqnarray}
where in the last step the $s$-orbit of the hydrogenic model is applied
for the in-plane wavefunction overlap. Obviously, increasing the Bohr
radius $a^{*}$ as well as reducing the overlap of the out-of-plane wavefunction between the electron and hole contribute to longer lifetime of the exciton.

For a system without $\vec{E}_b$, such as in $\mathrm{MoS_{2}}$ monolayer, the screening effect is weak, and the exciton binds strongly, which results in relatively narrow exciton Bohr radius ($a^{*}$) and high binding energy ($E_{b}$)(see \autoref{tab:table1}). Consequently, the electron-hole pair is hard to separate. In addition, as depicted in \autoref{fig:demo}(a), the center of the out-of-plane wavefunctions $\varphi_{c\left(v\right)}\left(z\right)$ are coincident, which give rise to relatively large out-of-plane wavefunction overlap, resulting in fast recombination rate. By contrast, the screening effect becomes much stronger if a large $\vec{E}_b$ is present, for example in $\mathrm{MoSTe}$ monolayer. In this sense, the exciton binds loosely, giving rise to a relatively wide spatial extension and a low exciton binding energy. As a result, a high probability of successful separation for electron-hole pairs can be expected. Meanwhile, the strong build-in electric field also separates the center of the out-of-plane wavefunctions  (see \autoref{fig:demo}(b)), which significantly reduces the overlap of the out-of-plane wavefunctions, leading to long lifetime of the exciton.

\begin{figure}[t]
  \centering
\includegraphics[width=3in]{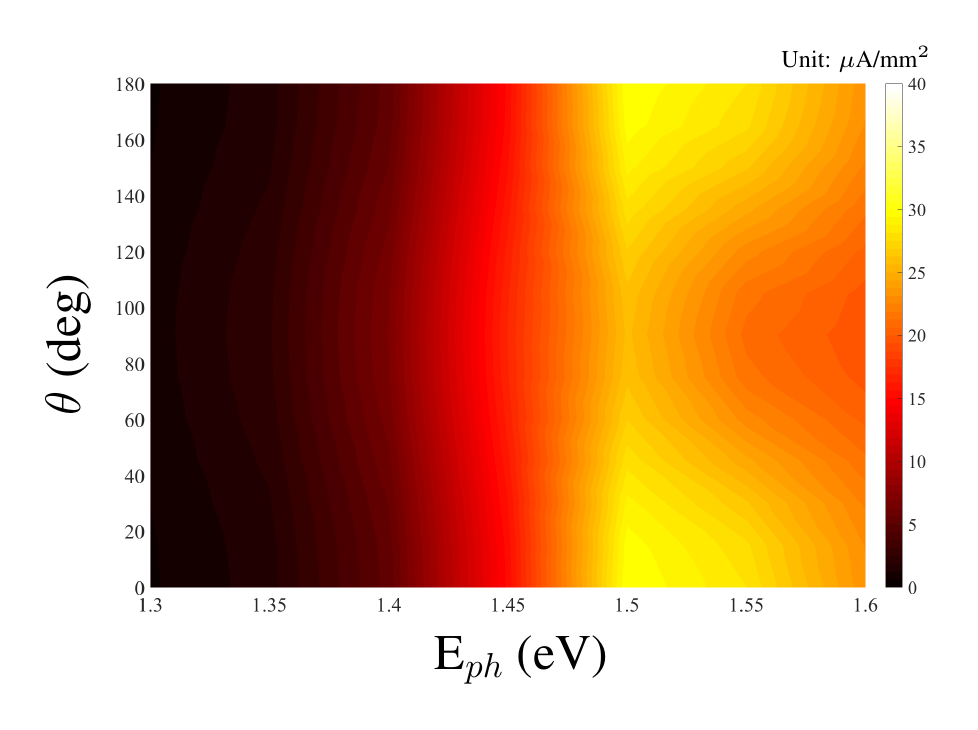}
\caption{Simulated photocurrents $I_{ph}$ as a function of photon energy and polarizing angle ($\theta$) for MoSTe monolayer.}
\label{fig:Iph}
\end{figure}

In this work, we also evaluate the photoresponse performance of MoSTe within the framework of NEGF-DFT. Upon illumination, electrons-hole pairs are created and traverse the device from left to the right sides, leading to the formation of photoinduced current throughout the monolayer. In our simulations, linearly polarized light which propagates perpendicular to the janus-MoSTe monolayer is used to irradiate the system. The incident light power density is l kW/m, i.e. AM1.5, which is the typical value of the solar power. The predicted photocurrents for janus-MoSTe at different polarized angle $\theta$ with photon energies ranging from 1.3 to 1.6 eV are shown in \autoref{fig:Iph}. From the results, we find that in consistent with our previous analysis, janus-MoSTe monolayer shows high photoresponsivity with the photocurrent up to 30 $\mu$A/mm$^2$, suggesting powerful potentials in further  optoelectronic and photovoltaic applications.

\section{Conclusions}
In summary, by employing TD-DFT combined with NAMD, we systematically studied the electronic properties of monolayer janus-MoSTe. Our results show that due to the difference of electronegativity between S and Te atoms, janus-MoSTe exhibits an out-of-plane build-in electric field, which significantly reduces the $e-h$ interaction. In this sense, the exciton binds loosely, giving rise to a relatively wide spatial extension and a low exciton binding energy. The calculated exciton binding energy is as low as 0.54 eV according to GW-BSE approach. The electron-hole recombination rate is then evaluated based on the nonadiabatic molecular dynamics (NAMD) technique. Our calculations predict that the present of the build-in electric filed substantially reduces the nonradiative recombination rate, with the lifetime up to 1.31 ns, which is even comparable with those of van der Waals (vdW) heterostructures made by stacking layered TMDs. Such extremely long exciton lifetime is interpreted by the strong Coulomb screening effect as well as the the greatly reduced overlap of the out-of-plane wavefunctions between the electron and hole, which both result from the large build-in electric field. In addition, the photoresponse performance of janus-MoSTe is also evaluated within the framework of NEGF-DFT. Upon illumination, janus-MoSTe monolayer shows high photoresponsivity with the photocurrent up to 30 $\mu$A/mm$^2$. The extremely long timescale for electron-hole recombination combined with high photoresponse performance make janus-MoSTe monolayer promising candidate for further applications in light detecting and harvesting.

\section*{Acknowledgment}
This work is supported by the National Natural Science Foundation of China (No.11604213, and No.11574217), Shenzhen Key Lab Fund (Grant No. ZDSYS20170228105421966), and Taishan Scholar Program of Shandong Provice.

%


\begin{thebibliography}{39}%
\makeatletter
\providecommand \@ifxundefined [1]{%
 \@ifx{#1\undefined}
}%
\providecommand \@ifnum [1]{%
 \ifnum #1\expandafter \@firstoftwo
 \else \expandafter \@secondoftwo
 \fi
}%
\providecommand \@ifx [1]{%
 \ifx #1\expandafter \@firstoftwo
 \else \expandafter \@secondoftwo
 \fi
}%
\providecommand \natexlab [1]{#1}%
\providecommand \enquote  [1]{``#1''}%
\providecommand \bibnamefont  [1]{#1}%
\providecommand \bibfnamefont [1]{#1}%
\providecommand \citenamefont [1]{#1}%
\providecommand \href@noop [0]{\@secondoftwo}%
\providecommand \href [0]{\begingroup \@sanitize@url \@href}%
\providecommand \@href[1]{\@@startlink{#1}\@@href}%
\providecommand \@@href[1]{\endgroup#1\@@endlink}%
\providecommand \@sanitize@url [0]{\catcode `\\12\catcode `\$12\catcode
  `\&12\catcode `\#12\catcode `\^12\catcode `\_12\catcode `\%12\relax}%
\providecommand \@@startlink[1]{}%
\providecommand \@@endlink[0]{}%
\providecommand \url  [0]{\begingroup\@sanitize@url \@url }%
\providecommand \@url [1]{\endgroup\@href {#1}{\urlprefix }}%
\providecommand \urlprefix  [0]{URL }%
\providecommand \Eprint [0]{\href }%
\providecommand \doibase [0]{http://dx.doi.org/}%
\providecommand \selectlanguage [0]{\@gobble}%
\providecommand \bibinfo  [0]{\@secondoftwo}%
\providecommand \bibfield  [0]{\@secondoftwo}%
\providecommand \translation [1]{[#1]}%
\providecommand \BibitemOpen [0]{}%
\providecommand \bibitemStop [0]{}%
\providecommand \bibitemNoStop [0]{.\EOS\space}%
\providecommand \EOS [0]{\spacefactor3000\relax}%
\providecommand \BibitemShut  [1]{\csname bibitem#1\endcsname}%
\let\auto@bib@innerbib\@empty
\bibitem [{\citenamefont {Novoselov}\ \emph {et~al.}(2004)\citenamefont
  {Novoselov}, \citenamefont {Geim}, \citenamefont {Morozov}, \citenamefont
  {Jiang}, \citenamefont {Zhang}, \citenamefont {Dubonos}, \citenamefont
  {Grigorieva},\ and\ \citenamefont {Firsov}}]{RN724}%
  \BibitemOpen
  \bibfield  {author} {\bibinfo {author} {\bibfnamefont {K.~S.}\ \bibnamefont
  {Novoselov}}, \bibinfo {author} {\bibfnamefont {A.~K.}\ \bibnamefont {Geim}},
  \bibinfo {author} {\bibfnamefont {S.~V.}\ \bibnamefont {Morozov}}, \bibinfo
  {author} {\bibfnamefont {D.}~\bibnamefont {Jiang}}, \bibinfo {author}
  {\bibfnamefont {Y.}~\bibnamefont {Zhang}}, \bibinfo {author} {\bibfnamefont
  {S.~V.}\ \bibnamefont {Dubonos}}, \bibinfo {author} {\bibfnamefont {I.~V.}\
  \bibnamefont {Grigorieva}}, \ and\ \bibinfo {author} {\bibfnamefont {A.~A.}\
  \bibnamefont {Firsov}},\ }\href@noop {} {\bibfield  {journal} {\bibinfo
  {journal} {Science}\ }\textbf {\bibinfo {volume} {306}},\ \bibinfo {pages}
  {666} (\bibinfo {year} {2004})}\BibitemShut {NoStop}%
\bibitem [{\citenamefont {Hernandez}\ \emph {et~al.}(2008)\citenamefont
  {Hernandez}, \citenamefont {Nicolosi}, \citenamefont {Lotya}, \citenamefont
  {Blighe}, \citenamefont {Sun}, \citenamefont {De}, \citenamefont {McGovern},
  \citenamefont {Holland}, \citenamefont {Byrne}, \citenamefont {Gun'Ko},
  \citenamefont {Boland}, \citenamefont {Niraj}, \citenamefont {Duesberg},
  \citenamefont {Krishnamurthy}, \citenamefont {Goodhue}, \citenamefont
  {Hutchison}, \citenamefont {Scardaci}, \citenamefont {Ferrari},\ and\
  \citenamefont {Coleman}}]{hernandez2008high}%
  \BibitemOpen
  \bibfield  {author} {\bibinfo {author} {\bibfnamefont {Y.}~\bibnamefont
  {Hernandez}}, \bibinfo {author} {\bibfnamefont {V.}~\bibnamefont {Nicolosi}},
  \bibinfo {author} {\bibfnamefont {M.}~\bibnamefont {Lotya}}, \bibinfo
  {author} {\bibfnamefont {F.~M.}\ \bibnamefont {Blighe}}, \bibinfo {author}
  {\bibfnamefont {Z.}~\bibnamefont {Sun}}, \bibinfo {author} {\bibfnamefont
  {S.}~\bibnamefont {De}}, \bibinfo {author} {\bibfnamefont {I.}~\bibnamefont
  {McGovern}}, \bibinfo {author} {\bibfnamefont {B.}~\bibnamefont {Holland}},
  \bibinfo {author} {\bibfnamefont {M.}~\bibnamefont {Byrne}}, \bibinfo
  {author} {\bibfnamefont {Y.~K.}\ \bibnamefont {Gun'Ko}}, \bibinfo {author}
  {\bibfnamefont {J.~J.}\ \bibnamefont {Boland}}, \bibinfo {author}
  {\bibfnamefont {P.}~\bibnamefont {Niraj}}, \bibinfo {author} {\bibfnamefont
  {G.}~\bibnamefont {Duesberg}}, \bibinfo {author} {\bibfnamefont
  {S.}~\bibnamefont {Krishnamurthy}}, \bibinfo {author} {\bibfnamefont
  {R.}~\bibnamefont {Goodhue}}, \bibinfo {author} {\bibfnamefont
  {J.}~\bibnamefont {Hutchison}}, \bibinfo {author} {\bibfnamefont
  {V.}~\bibnamefont {Scardaci}}, \bibinfo {author} {\bibfnamefont {A.~C.}\
  \bibnamefont {Ferrari}}, \ and\ \bibinfo {author} {\bibfnamefont {J.~N.}\
  \bibnamefont {Coleman}},\ }\href@noop {} {\bibfield  {journal} {\bibinfo
  {journal} {Nature Nanotechnol.}\ }\textbf {\bibinfo {volume} {3}},\ \bibinfo
  {pages} {563} (\bibinfo {year} {2008})}\BibitemShut {NoStop}%
\bibitem [{\citenamefont {Mak}\ \emph {et~al.}(2010)\citenamefont {Mak},
  \citenamefont {Lee}, \citenamefont {Hone}, \citenamefont {Shan},\ and\
  \citenamefont {Heinz}}]{Mark10}%
  \BibitemOpen
  \bibfield  {author} {\bibinfo {author} {\bibfnamefont {K.~F.}\ \bibnamefont
  {Mak}}, \bibinfo {author} {\bibfnamefont {C.}~\bibnamefont {Lee}}, \bibinfo
  {author} {\bibfnamefont {J.}~\bibnamefont {Hone}}, \bibinfo {author}
  {\bibfnamefont {J.}~\bibnamefont {Shan}}, \ and\ \bibinfo {author}
  {\bibfnamefont {T.~F.}\ \bibnamefont {Heinz}},\ }\href {\doibase
  10.1103/PhysRevLett.105.136805} {\bibfield  {journal} {\bibinfo  {journal}
  {Phys. Rev. Lett.}\ }\textbf {\bibinfo {volume} {105}},\ \bibinfo {pages}
  {136805} (\bibinfo {year} {2010})}\BibitemShut {NoStop}%
\bibitem [{\citenamefont {Wang}\ \emph {et~al.}(2012)\citenamefont {Wang},
  \citenamefont {Kalantar-Zadeh}, \citenamefont {Kis}, \citenamefont
  {Coleman},\ and\ \citenamefont {Strano}}]{MoS212}%
  \BibitemOpen
  \bibfield  {author} {\bibinfo {author} {\bibfnamefont {Q.~H.}\ \bibnamefont
  {Wang}}, \bibinfo {author} {\bibfnamefont {K.}~\bibnamefont
  {Kalantar-Zadeh}}, \bibinfo {author} {\bibfnamefont {A.}~\bibnamefont {Kis}},
  \bibinfo {author} {\bibfnamefont {J.~N.}\ \bibnamefont {Coleman}}, \ and\
  \bibinfo {author} {\bibfnamefont {M.~S.}\ \bibnamefont {Strano}},\ }\href
  {\doibase 10.1038/NNANO.2012.193} {\bibfield  {journal} {\bibinfo  {journal}
  {Nature Nanotechnol.}\ }\textbf {\bibinfo {volume} {7}},\ \bibinfo {pages}
  {699} (\bibinfo {year} {2012})}\BibitemShut {NoStop}%
\bibitem [{\citenamefont {Xia}\ \emph {et~al.}(2014)\citenamefont {Xia},
  \citenamefont {Wang},\ and\ \citenamefont {Jia}}]{xia2014rediscovering}%
  \BibitemOpen
  \bibfield  {author} {\bibinfo {author} {\bibfnamefont {F.}~\bibnamefont
  {Xia}}, \bibinfo {author} {\bibfnamefont {H.}~\bibnamefont {Wang}}, \ and\
  \bibinfo {author} {\bibfnamefont {Y.}~\bibnamefont {Jia}},\ }\href@noop {}
  {\bibfield  {journal} {\bibinfo  {journal} {Nat. Commun.}\ }\textbf {\bibinfo
  {volume} {5}},\ \bibinfo {pages} {4458} (\bibinfo {year} {2014})}\BibitemShut
  {NoStop}%
\bibitem [{\citenamefont {Ye}\ \emph {et~al.}(2014)\citenamefont {Ye},
  \citenamefont {Cao}, \citenamefont {O��Brien}, \citenamefont {Zhu},
  \citenamefont {Yin}, \citenamefont {Wang}, \citenamefont {Louie},\ and\
  \citenamefont {Zhang}}]{Ye2014Probing}%
  \BibitemOpen
  \bibfield  {author} {\bibinfo {author} {\bibfnamefont {Z.}~\bibnamefont
  {Ye}}, \bibinfo {author} {\bibfnamefont {T.}~\bibnamefont {Cao}}, \bibinfo
  {author} {\bibfnamefont {K.}~\bibnamefont {O��Brien}}, \bibinfo {author}
  {\bibfnamefont {H.}~\bibnamefont {Zhu}}, \bibinfo {author} {\bibfnamefont
  {X.}~\bibnamefont {Yin}}, \bibinfo {author} {\bibfnamefont {Y.}~\bibnamefont
  {Wang}}, \bibinfo {author} {\bibfnamefont {S.~G.}\ \bibnamefont {Louie}}, \
  and\ \bibinfo {author} {\bibfnamefont {X.}~\bibnamefont {Zhang}},\
  }\href@noop {} {\bibfield  {journal} {\bibinfo  {journal} {Nature}\ }\textbf
  {\bibinfo {volume} {513}},\ \bibinfo {pages} {214} (\bibinfo {year}
  {2014})}\BibitemShut {NoStop}%
\bibitem [{\citenamefont {Peng}\ \emph {et~al.}(2015)\citenamefont {Peng},
  \citenamefont {Ang},\ and\ \citenamefont {Loh}}]{PENG2015128}%
  \BibitemOpen
  \bibfield  {author} {\bibinfo {author} {\bibfnamefont {B.}~\bibnamefont
  {Peng}}, \bibinfo {author} {\bibfnamefont {P.~K.}\ \bibnamefont {Ang}}, \
  and\ \bibinfo {author} {\bibfnamefont {K.~P.}\ \bibnamefont {Loh}},\ }\href
  {\doibase https://doi.org/10.1016/j.nantod.2015.01.007} {\bibfield  {journal}
  {\bibinfo  {journal} {Nano Today}\ }\textbf {\bibinfo {volume} {10}},\
  \bibinfo {pages} {128 } (\bibinfo {year} {2015})}\BibitemShut {NoStop}%
\bibitem [{\citenamefont {Korn}\ \emph {et~al.}(2011)\citenamefont {Korn},
  \citenamefont {Heydrich}, \citenamefont {Hirmer}, \citenamefont
  {Schmutzler},\ and\ \citenamefont {Sch��ller}}]{Korn2011}%
  \BibitemOpen
  \bibfield  {author} {\bibinfo {author} {\bibfnamefont {T.}~\bibnamefont
  {Korn}}, \bibinfo {author} {\bibfnamefont {S.}~\bibnamefont {Heydrich}},
  \bibinfo {author} {\bibfnamefont {M.}~\bibnamefont {Hirmer}}, \bibinfo
  {author} {\bibfnamefont {J.}~\bibnamefont {Schmutzler}}, \ and\ \bibinfo
  {author} {\bibfnamefont {C.}~\bibnamefont {Sch��ller}},\ }\href {\doibase
  10.1063/1.3636402} {\bibfield  {journal} {\bibinfo  {journal} {Appl. Phys.
  Lett.}\ }\textbf {\bibinfo {volume} {99}},\ \bibinfo {pages} {102109}
  (\bibinfo {year} {2011})}\BibitemShut {NoStop}%
\bibitem [{\citenamefont {Sun}\ \emph {et~al.}(2014)\citenamefont {Sun},
  \citenamefont {Rao}, \citenamefont {Reider}, \citenamefont {Chen},
  \citenamefont {You}, \citenamefont {Br��zin}, \citenamefont {Harutyunyan},\
  and\ \citenamefont {Heinz}}]{Sun2014}%
  \BibitemOpen
  \bibfield  {author} {\bibinfo {author} {\bibfnamefont {D.}~\bibnamefont
  {Sun}}, \bibinfo {author} {\bibfnamefont {Y.}~\bibnamefont {Rao}}, \bibinfo
  {author} {\bibfnamefont {G.~A.}\ \bibnamefont {Reider}}, \bibinfo {author}
  {\bibfnamefont {G.}~\bibnamefont {Chen}}, \bibinfo {author} {\bibfnamefont
  {Y.}~\bibnamefont {You}}, \bibinfo {author} {\bibfnamefont {L.}~\bibnamefont
  {Br��zin}}, \bibinfo {author} {\bibfnamefont {A.~R.}\ \bibnamefont
  {Harutyunyan}}, \ and\ \bibinfo {author} {\bibfnamefont {T.~F.}\ \bibnamefont
  {Heinz}},\ }\href {\doibase 10.1021/nl5021975} {\bibfield  {journal}
  {\bibinfo  {journal} {Nano Lett.}\ }\textbf {\bibinfo {volume} {14}},\
  \bibinfo {pages} {5625} (\bibinfo {year} {2014})}\BibitemShut {NoStop}%
\bibitem [{\citenamefont {Li}\ \emph {et~al.}(2017{\natexlab{a}})\citenamefont
  {Li}, \citenamefont {Long}, \citenamefont {Bertolini},\ and\ \citenamefont
  {Prezhdo}}]{Linqiu17}%
  \BibitemOpen
  \bibfield  {author} {\bibinfo {author} {\bibfnamefont {L.}~\bibnamefont
  {Li}}, \bibinfo {author} {\bibfnamefont {R.}~\bibnamefont {Long}}, \bibinfo
  {author} {\bibfnamefont {T.}~\bibnamefont {Bertolini}}, \ and\ \bibinfo
  {author} {\bibfnamefont {O.~V.}\ \bibnamefont {Prezhdo}},\ }\href {\doibase
  10.1021/acs.nanolett.7b04374} {\bibfield  {journal} {\bibinfo  {journal}
  {Nano Lett.}\ }\textbf {\bibinfo {volume} {17}},\ \bibinfo {pages} {7962}
  (\bibinfo {year} {2017}{\natexlab{a}})}\BibitemShut {NoStop}%
\bibitem [{\citenamefont {Sundaram}\ \emph {et~al.}(2013)\citenamefont
  {Sundaram}, \citenamefont {Engel}, \citenamefont {Lombardo}, \citenamefont
  {Krupke}, \citenamefont {Ferrari}, \citenamefont {Avouris},\ and\
  \citenamefont {Steiner}}]{sundaram2013electroluminescence}%
  \BibitemOpen
  \bibfield  {author} {\bibinfo {author} {\bibfnamefont {R.}~\bibnamefont
  {Sundaram}}, \bibinfo {author} {\bibfnamefont {M.}~\bibnamefont {Engel}},
  \bibinfo {author} {\bibfnamefont {A.}~\bibnamefont {Lombardo}}, \bibinfo
  {author} {\bibfnamefont {R.}~\bibnamefont {Krupke}}, \bibinfo {author}
  {\bibfnamefont {A.}~\bibnamefont {Ferrari}}, \bibinfo {author} {\bibfnamefont
  {P.}~\bibnamefont {Avouris}}, \ and\ \bibinfo {author} {\bibfnamefont
  {M.}~\bibnamefont {Steiner}},\ }\href@noop {} {\bibfield  {journal} {\bibinfo
   {journal} {Nano Lett.}\ }\textbf {\bibinfo {volume} {13}},\ \bibinfo {pages}
  {1416} (\bibinfo {year} {2013})}\BibitemShut {NoStop}%
\bibitem [{\citenamefont {Ross}\ \emph {et~al.}(2014)\citenamefont {Ross},
  \citenamefont {Klement}, \citenamefont {Jones}, \citenamefont {Ghimire},
  \citenamefont {Yan}, \citenamefont {Mandrus}, \citenamefont {Taniguchi},
  \citenamefont {Watanabe}, \citenamefont {Kitamura},\ and\ \citenamefont
  {Yao}}]{ross2014electrically}%
  \BibitemOpen
  \bibfield  {author} {\bibinfo {author} {\bibfnamefont {J.~S.}\ \bibnamefont
  {Ross}}, \bibinfo {author} {\bibfnamefont {P.}~\bibnamefont {Klement}},
  \bibinfo {author} {\bibfnamefont {A.~M.}\ \bibnamefont {Jones}}, \bibinfo
  {author} {\bibfnamefont {N.~J.}\ \bibnamefont {Ghimire}}, \bibinfo {author}
  {\bibfnamefont {J.}~\bibnamefont {Yan}}, \bibinfo {author} {\bibfnamefont
  {D.~G.}\ \bibnamefont {Mandrus}}, \bibinfo {author} {\bibfnamefont
  {T.}~\bibnamefont {Taniguchi}}, \bibinfo {author} {\bibfnamefont
  {K.}~\bibnamefont {Watanabe}}, \bibinfo {author} {\bibfnamefont
  {K.}~\bibnamefont {Kitamura}}, \ and\ \bibinfo {author} {\bibfnamefont
  {W.}~\bibnamefont {Yao}},\ }\href@noop {} {\bibfield  {journal} {\bibinfo
  {journal} {Nature Nanotechnol.}\ }\textbf {\bibinfo {volume} {9}},\ \bibinfo
  {pages} {268} (\bibinfo {year} {2014})}\BibitemShut {NoStop}%
\bibitem [{\citenamefont {Wang}\ \emph {et~al.}(2015)\citenamefont {Wang},
  \citenamefont {Zhang}, \citenamefont {Chan}, \citenamefont {Tiwari},\ and\
  \citenamefont {Rana}}]{Wang2015Ultrafast}%
  \BibitemOpen
  \bibfield  {author} {\bibinfo {author} {\bibfnamefont {H.}~\bibnamefont
  {Wang}}, \bibinfo {author} {\bibfnamefont {C.}~\bibnamefont {Zhang}},
  \bibinfo {author} {\bibfnamefont {W.}~\bibnamefont {Chan}}, \bibinfo {author}
  {\bibfnamefont {S.}~\bibnamefont {Tiwari}}, \ and\ \bibinfo {author}
  {\bibfnamefont {F.}~\bibnamefont {Rana}},\ }\href@noop {} {\bibfield
  {journal} {\bibinfo  {journal} {Nat. Commun.}\ }\textbf {\bibinfo {volume}
  {6}},\ \bibinfo {pages} {8831} (\bibinfo {year} {2015})}\BibitemShut
  {NoStop}%
\bibitem [{\citenamefont {Lu}\ \emph {et~al.}(2017)\citenamefont {Lu},
  \citenamefont {Zhu}, \citenamefont {Xiao}, \citenamefont {Chuu},
  \citenamefont {Han}, \citenamefont {Chiu}, \citenamefont {Cheng},
  \citenamefont {Yang}, \citenamefont {Wei}, \citenamefont {Yang},
  \citenamefont {Wang}, \citenamefont {Sokaras}, \citenamefont {Nordlund},
  \citenamefont {Yang}, \citenamefont {Muller}, \citenamefont {Chou},
  \citenamefont {Zhang},\ and\ \citenamefont {Li}}]{lu2017janus}%
  \BibitemOpen
  \bibfield  {author} {\bibinfo {author} {\bibfnamefont {A.-Y.}\ \bibnamefont
  {Lu}}, \bibinfo {author} {\bibfnamefont {H.}~\bibnamefont {Zhu}}, \bibinfo
  {author} {\bibfnamefont {J.}~\bibnamefont {Xiao}}, \bibinfo {author}
  {\bibfnamefont {C.-P.}\ \bibnamefont {Chuu}}, \bibinfo {author}
  {\bibfnamefont {Y.}~\bibnamefont {Han}}, \bibinfo {author} {\bibfnamefont
  {M.-H.}\ \bibnamefont {Chiu}}, \bibinfo {author} {\bibfnamefont {C.-C.}\
  \bibnamefont {Cheng}}, \bibinfo {author} {\bibfnamefont {C.-W.}\ \bibnamefont
  {Yang}}, \bibinfo {author} {\bibfnamefont {K.-H.}\ \bibnamefont {Wei}},
  \bibinfo {author} {\bibfnamefont {Y.}~\bibnamefont {Yang}}, \bibinfo {author}
  {\bibfnamefont {Y.}~\bibnamefont {Wang}}, \bibinfo {author} {\bibfnamefont
  {D.}~\bibnamefont {Sokaras}}, \bibinfo {author} {\bibfnamefont
  {D.}~\bibnamefont {Nordlund}}, \bibinfo {author} {\bibfnamefont
  {P.}~\bibnamefont {Yang}}, \bibinfo {author} {\bibfnamefont {D.~A.}\
  \bibnamefont {Muller}}, \bibinfo {author} {\bibfnamefont {M.-Y.}\
  \bibnamefont {Chou}}, \bibinfo {author} {\bibfnamefont {X.}~\bibnamefont
  {Zhang}}, \ and\ \bibinfo {author} {\bibfnamefont {L.-J.}\ \bibnamefont
  {Li}},\ }\href@noop {} {\bibfield  {journal} {\bibinfo  {journal} {Nature
  Nanotechnol.}\ }\textbf {\bibinfo {volume} {12}},\ \bibinfo {pages} {744}
  (\bibinfo {year} {2017})}\BibitemShut {NoStop}%
\bibitem [{\citenamefont {Zhang}\ \emph {et~al.}(2017)\citenamefont {Zhang},
  \citenamefont {Jia}, \citenamefont {Kholmanov}, \citenamefont {Dong},
  \citenamefont {Er}, \citenamefont {Chen}, \citenamefont {Guo}, \citenamefont
  {Jin}, \citenamefont {Shenoy}, \citenamefont {Shi},\ and\ \citenamefont
  {Lou}}]{zhang2017janus}%
  \BibitemOpen
  \bibfield  {author} {\bibinfo {author} {\bibfnamefont {J.}~\bibnamefont
  {Zhang}}, \bibinfo {author} {\bibfnamefont {S.}~\bibnamefont {Jia}}, \bibinfo
  {author} {\bibfnamefont {I.}~\bibnamefont {Kholmanov}}, \bibinfo {author}
  {\bibfnamefont {L.}~\bibnamefont {Dong}}, \bibinfo {author} {\bibfnamefont
  {D.}~\bibnamefont {Er}}, \bibinfo {author} {\bibfnamefont {W.}~\bibnamefont
  {Chen}}, \bibinfo {author} {\bibfnamefont {H.}~\bibnamefont {Guo}}, \bibinfo
  {author} {\bibfnamefont {Z.}~\bibnamefont {Jin}}, \bibinfo {author}
  {\bibfnamefont {V.~B.}\ \bibnamefont {Shenoy}}, \bibinfo {author}
  {\bibfnamefont {L.}~\bibnamefont {Shi}}, \ and\ \bibinfo {author}
  {\bibfnamefont {J.}~\bibnamefont {Lou}},\ }\href@noop {} {\bibfield
  {journal} {\bibinfo  {journal} {ACS Nano}\ }\textbf {\bibinfo {volume}
  {11}},\ \bibinfo {pages} {8192} (\bibinfo {year} {2017})}\BibitemShut
  {NoStop}%
\bibitem [{\citenamefont {Li}\ \emph {et~al.}(2017{\natexlab{b}})\citenamefont
  {Li}, \citenamefont {Wei}, \citenamefont {Zhao}, \citenamefont {Huang},\ and\
  \citenamefont {Dai}}]{li2017electronic}%
  \BibitemOpen
  \bibfield  {author} {\bibinfo {author} {\bibfnamefont {F.}~\bibnamefont
  {Li}}, \bibinfo {author} {\bibfnamefont {W.}~\bibnamefont {Wei}}, \bibinfo
  {author} {\bibfnamefont {P.}~\bibnamefont {Zhao}}, \bibinfo {author}
  {\bibfnamefont {B.}~\bibnamefont {Huang}}, \ and\ \bibinfo {author}
  {\bibfnamefont {Y.}~\bibnamefont {Dai}},\ }\href@noop {} {\bibfield
  {journal} {\bibinfo  {journal} {J. Phys. Chem. Lett.}\ }\textbf {\bibinfo
  {volume} {8}},\ \bibinfo {pages} {5959} (\bibinfo {year}
  {2017}{\natexlab{b}})}\BibitemShut {NoStop}%
\bibitem [{\citenamefont {Rivera}\ \emph {et~al.}(2015)\citenamefont {Rivera},
  \citenamefont {Schaibley}, \citenamefont {Jones}, \citenamefont {Ross},
  \citenamefont {Wu}, \citenamefont {Aivazian}, \citenamefont {Klement},
  \citenamefont {Seyler}, \citenamefont {Clark},\ and\ \citenamefont
  {Ghimire}}]{Rivera2015Observation}%
  \BibitemOpen
  \bibfield  {author} {\bibinfo {author} {\bibfnamefont {P.}~\bibnamefont
  {Rivera}}, \bibinfo {author} {\bibfnamefont {J.~R.}\ \bibnamefont
  {Schaibley}}, \bibinfo {author} {\bibfnamefont {A.~M.}\ \bibnamefont
  {Jones}}, \bibinfo {author} {\bibfnamefont {J.~S.}\ \bibnamefont {Ross}},
  \bibinfo {author} {\bibfnamefont {S.}~\bibnamefont {Wu}}, \bibinfo {author}
  {\bibfnamefont {G.}~\bibnamefont {Aivazian}}, \bibinfo {author}
  {\bibfnamefont {P.}~\bibnamefont {Klement}}, \bibinfo {author} {\bibfnamefont
  {K.}~\bibnamefont {Seyler}}, \bibinfo {author} {\bibfnamefont
  {G.}~\bibnamefont {Clark}}, \ and\ \bibinfo {author} {\bibfnamefont {N.~J.}\
  \bibnamefont {Ghimire}},\ }\href@noop {} {\bibfield  {journal} {\bibinfo
  {journal} {Nat. Commun.}\ }\textbf {\bibinfo {volume} {6}},\ \bibinfo {pages}
  {6242} (\bibinfo {year} {2015})}\BibitemShut {NoStop}%
\bibitem [{\citenamefont {Li}\ \emph {et~al.}(2016)\citenamefont {Li},
  \citenamefont {Long},\ and\ \citenamefont {Prezhdo}}]{Li2016Charge}%
  \BibitemOpen
  \bibfield  {author} {\bibinfo {author} {\bibfnamefont {L.}~\bibnamefont
  {Li}}, \bibinfo {author} {\bibfnamefont {R.}~\bibnamefont {Long}}, \ and\
  \bibinfo {author} {\bibfnamefont {O.~V.}\ \bibnamefont {Prezhdo}},\
  }\href@noop {} {\bibfield  {journal} {\bibinfo  {journal} {Chem. Mater.}\
  }\textbf {\bibinfo {volume} {29}},\ \bibinfo {pages} {2466} (\bibinfo {year}
  {2016})}\BibitemShut {NoStop}%
\bibitem [{\citenamefont {Kresse}\ and\ \citenamefont
  {Joubert}(1999)}]{kresse1999ultrasoft}%
  \BibitemOpen
  \bibfield  {author} {\bibinfo {author} {\bibfnamefont {G.}~\bibnamefont
  {Kresse}}\ and\ \bibinfo {author} {\bibfnamefont {D.}~\bibnamefont
  {Joubert}},\ }\href@noop {} {\bibfield  {journal} {\bibinfo  {journal} {Phys.
  Rev. B}\ }\textbf {\bibinfo {volume} {59}},\ \bibinfo {pages} {1758}
  (\bibinfo {year} {1999})}\BibitemShut {NoStop}%
\bibitem [{\citenamefont {Kresse}\ and\ \citenamefont
  {Furthm\"{u}ller}(1996{\natexlab{a}})}]{KressePRB96}%
  \BibitemOpen
  \bibfield  {author} {\bibinfo {author} {\bibfnamefont {G.}~\bibnamefont
  {Kresse}}\ and\ \bibinfo {author} {\bibfnamefont {J.}~\bibnamefont
  {Furthm\"{u}ller}},\ }\href@noop {} {\bibfield  {journal} {\bibinfo
  {journal} {Phys. Rev. B}\ }\textbf {\bibinfo {volume} {54}},\ \bibinfo
  {pages} {11169} (\bibinfo {year} {1996}{\natexlab{a}})}\BibitemShut {NoStop}%
\bibitem [{\citenamefont {Kresse}\ and\ \citenamefont
  {Furthm\"{u}ller}(1996{\natexlab{b}})}]{Kresse96}%
  \BibitemOpen
  \bibfield  {author} {\bibinfo {author} {\bibfnamefont {G.}~\bibnamefont
  {Kresse}}\ and\ \bibinfo {author} {\bibfnamefont {J.}~\bibnamefont
  {Furthm\"{u}ller}},\ }\href@noop {} {\bibfield  {journal} {\bibinfo
  {journal} {Comput. Mater. Sci.}\ }\textbf {\bibinfo {volume} {6}},\ \bibinfo
  {pages} {15} (\bibinfo {year} {1996}{\natexlab{b}})}\BibitemShut {NoStop}%
\bibitem [{\citenamefont {Perdew}\ \emph {et~al.}(1996)\citenamefont {Perdew},
  \citenamefont {Burke},\ and\ \citenamefont
  {Ernzerhof}}]{perdew1996generalized}%
  \BibitemOpen
  \bibfield  {author} {\bibinfo {author} {\bibfnamefont {J.~P.}\ \bibnamefont
  {Perdew}}, \bibinfo {author} {\bibfnamefont {K.}~\bibnamefont {Burke}}, \
  and\ \bibinfo {author} {\bibfnamefont {M.}~\bibnamefont {Ernzerhof}},\
  }\href@noop {} {\bibfield  {journal} {\bibinfo  {journal} {Phys. Rev. lett.}\
  }\textbf {\bibinfo {volume} {77}},\ \bibinfo {pages} {3865} (\bibinfo {year}
  {1996})}\BibitemShut {NoStop}%
\bibitem [{\citenamefont {Grimme}(2006)}]{Grimme06}%
  \BibitemOpen
  \bibfield  {author} {\bibinfo {author} {\bibfnamefont {S.}~\bibnamefont
  {Grimme}},\ }\href {\doibase 10.1002/jcc.20495} {\bibfield  {journal}
  {\bibinfo  {journal} {J. Comput. Chem.}\ }\textbf {\bibinfo {volume} {27}},\
  \bibinfo {pages} {1787} (\bibinfo {year} {2006})}\BibitemShut {NoStop}%
\bibitem [{\citenamefont {Heyd}\ \emph {et~al.}(2003)\citenamefont {Heyd},
  \citenamefont {Scuseria},\ and\ \citenamefont {Ernzerhof}}]{Heyd03}%
  \BibitemOpen
  \bibfield  {author} {\bibinfo {author} {\bibfnamefont {J.}~\bibnamefont
  {Heyd}}, \bibinfo {author} {\bibfnamefont {G.}~\bibnamefont {Scuseria}}, \
  and\ \bibinfo {author} {\bibfnamefont {M.}~\bibnamefont {Ernzerhof}},\ }\href
  {\doibase 10.1063/1.1564060} {\bibfield  {journal} {\bibinfo  {journal} {J.
  Chem. Phys.}\ }\textbf {\bibinfo {volume} {118}},\ \bibinfo {pages} {8207}
  (\bibinfo {year} {2003})}\BibitemShut {NoStop}%
\bibitem [{\citenamefont {Paier}\ \emph {et~al.}(2006)\citenamefont {Paier},
  \citenamefont {Marsman}, \citenamefont {Hummer}, \citenamefont {Kresse},
  \citenamefont {Gerber},\ and\ \citenamefont
  {$\acute{A}$ngy$\acute{a}$n}}]{Paier06}%
  \BibitemOpen
  \bibfield  {author} {\bibinfo {author} {\bibfnamefont {J.}~\bibnamefont
  {Paier}}, \bibinfo {author} {\bibfnamefont {M.}~\bibnamefont {Marsman}},
  \bibinfo {author} {\bibfnamefont {K.}~\bibnamefont {Hummer}}, \bibinfo
  {author} {\bibfnamefont {G.}~\bibnamefont {Kresse}}, \bibinfo {author}
  {\bibfnamefont {I.~C.}\ \bibnamefont {Gerber}}, \ and\ \bibinfo {author}
  {\bibfnamefont {J.~G.}\ \bibnamefont {$\acute{A}$ngy$\acute{a}$n}},\ }\href
  {\doibase http://dx.doi.org/10.1063/1.2187006} {\bibfield  {journal}
  {\bibinfo  {journal} {J. Chem. Phys.}\ }\textbf {\bibinfo {volume} {124}},\
  \bibinfo {pages} {154709} (\bibinfo {year} {2006})}\BibitemShut {NoStop}%
\bibitem [{\citenamefont {Rohlfing}\ and\ \citenamefont
  {Louie}(2000)}]{PhysRevB.62.4927}%
  \BibitemOpen
  \bibfield  {author} {\bibinfo {author} {\bibfnamefont {M.}~\bibnamefont
  {Rohlfing}}\ and\ \bibinfo {author} {\bibfnamefont {S.~G.}\ \bibnamefont
  {Louie}},\ }\href {\doibase 10.1103/PhysRevB.62.4927} {\bibfield  {journal}
  {\bibinfo  {journal} {Phys. Rev. B}\ }\textbf {\bibinfo {volume} {62}},\
  \bibinfo {pages} {4927} (\bibinfo {year} {2000})}\BibitemShut {NoStop}%
\bibitem [{\citenamefont {Shishkin}\ and\ \citenamefont
  {Kresse}(2006)}]{PhysRevB.74.035101}%
  \BibitemOpen
  \bibfield  {author} {\bibinfo {author} {\bibfnamefont {M.}~\bibnamefont
  {Shishkin}}\ and\ \bibinfo {author} {\bibfnamefont {G.}~\bibnamefont
  {Kresse}},\ }\href {\doibase 10.1103/PhysRevB.74.035101} {\bibfield
  {journal} {\bibinfo  {journal} {Phys. Rev. B}\ }\textbf {\bibinfo {volume}
  {74}},\ \bibinfo {pages} {035101} (\bibinfo {year} {2006})}\BibitemShut
  {NoStop}%
\bibitem [{\citenamefont {Olsen}\ \emph {et~al.}(2011)\citenamefont {Olsen},
  \citenamefont {Yan}, \citenamefont {Mortensen},\ and\ \citenamefont
  {Thygesen}}]{olsen2011dispersive}%
  \BibitemOpen
  \bibfield  {author} {\bibinfo {author} {\bibfnamefont {T.}~\bibnamefont
  {Olsen}}, \bibinfo {author} {\bibfnamefont {J.}~\bibnamefont {Yan}}, \bibinfo
  {author} {\bibfnamefont {J.~J.}\ \bibnamefont {Mortensen}}, \ and\ \bibinfo
  {author} {\bibfnamefont {K.~S.}\ \bibnamefont {Thygesen}},\ }\href@noop {}
  {\bibfield  {journal} {\bibinfo  {journal} {Phys. Rev. Lett.}\ }\textbf
  {\bibinfo {volume} {107}},\ \bibinfo {pages} {156401} (\bibinfo {year}
  {2011})}\BibitemShut {NoStop}%
\bibitem [{\citenamefont {Taylor}\ \emph {et~al.}(2001)\citenamefont {Taylor},
  \citenamefont {Guo},\ and\ \citenamefont {Wang}}]{nanodcal1}%
  \BibitemOpen
  \bibfield  {author} {\bibinfo {author} {\bibfnamefont {J.}~\bibnamefont
  {Taylor}}, \bibinfo {author} {\bibfnamefont {H.}~\bibnamefont {Guo}}, \ and\
  \bibinfo {author} {\bibfnamefont {J.}~\bibnamefont {Wang}},\ }\href {\doibase
  10.1103/PhysRevB.63.245407} {\bibfield  {journal} {\bibinfo  {journal} {Phys.
  Rev. B}\ }\textbf {\bibinfo {volume} {63}},\ \bibinfo {pages} {245407}
  (\bibinfo {year} {2001})}\BibitemShut {NoStop}%
\bibitem [{\citenamefont {Waldron}\ \emph {et~al.}(2006)\citenamefont
  {Waldron}, \citenamefont {Haney}, \citenamefont {Larade}, \citenamefont
  {MacDonald},\ and\ \citenamefont {Guo}}]{nanodcal2}%
  \BibitemOpen
  \bibfield  {author} {\bibinfo {author} {\bibfnamefont {D.}~\bibnamefont
  {Waldron}}, \bibinfo {author} {\bibfnamefont {P.}~\bibnamefont {Haney}},
  \bibinfo {author} {\bibfnamefont {B.}~\bibnamefont {Larade}}, \bibinfo
  {author} {\bibfnamefont {A.}~\bibnamefont {MacDonald}}, \ and\ \bibinfo
  {author} {\bibfnamefont {H.}~\bibnamefont {Guo}},\ }\href {\doibase
  10.1103/PhysRevLett.96.166804} {\bibfield  {journal} {\bibinfo  {journal}
  {Phys. Rev. Lett.}\ }\textbf {\bibinfo {volume} {96}},\ \bibinfo {pages}
  {166804} (\bibinfo {year} {2006})}\BibitemShut {NoStop}%
\bibitem [{\citenamefont {Haug}\ and\ \citenamefont
  {Jauho}(1998)}]{HaugQuantum}%
  \BibitemOpen
  \bibfield  {author} {\bibinfo {author} {\bibfnamefont {H.}~\bibnamefont
  {Haug}}\ and\ \bibinfo {author} {\bibfnamefont {A.~P.}\ \bibnamefont
  {Jauho}},\ }\href@noop {} {\emph {\bibinfo {title} {Quantum kinetics in
  transport and optics of semiconductors}}}\ (\bibinfo  {publisher}
  {Springer-Verlag, New York},\ \bibinfo {year} {1998})\ pp.\ \bibinfo {pages}
  {1077--1091}\BibitemShut {NoStop}%
\bibitem [{\citenamefont {Chen}\ \emph {et~al.}(2012)\citenamefont {Chen},
  \citenamefont {Hu},\ and\ \citenamefont {Guo}}]{Chen2012First}%
  \BibitemOpen
  \bibfield  {author} {\bibinfo {author} {\bibfnamefont {J.}~\bibnamefont
  {Chen}}, \bibinfo {author} {\bibfnamefont {Y.}~\bibnamefont {Hu}}, \ and\
  \bibinfo {author} {\bibfnamefont {H.}~\bibnamefont {Guo}},\ }\href@noop {}
  {\bibfield  {journal} {\bibinfo  {journal} {Phys. Rev. B}\ }\textbf {\bibinfo
  {volume} {341}},\ \bibinfo {pages} {324} (\bibinfo {year}
  {2012})}\BibitemShut {NoStop}%
\bibitem [{\citenamefont {Zhang}\ \emph {et~al.}(2014)\citenamefont {Zhang},
  \citenamefont {Gong}, \citenamefont {Chen}, \citenamefont {Liu},
  \citenamefont {Zhu}, \citenamefont {Xiao},\ and\ \citenamefont
  {Guo}}]{Zhang2014Generation}%
  \BibitemOpen
  \bibfield  {author} {\bibinfo {author} {\bibfnamefont {L.}~\bibnamefont
  {Zhang}}, \bibinfo {author} {\bibfnamefont {K.}~\bibnamefont {Gong}},
  \bibinfo {author} {\bibfnamefont {J.}~\bibnamefont {Chen}}, \bibinfo {author}
  {\bibfnamefont {L.}~\bibnamefont {Liu}}, \bibinfo {author} {\bibfnamefont
  {Y.}~\bibnamefont {Zhu}}, \bibinfo {author} {\bibfnamefont {D.}~\bibnamefont
  {Xiao}}, \ and\ \bibinfo {author} {\bibfnamefont {H.}~\bibnamefont {Guo}},\
  }\href@noop {} {\bibfield  {journal} {\bibinfo  {journal} {Phys. Rev. B}\
  }\textbf {\bibinfo {volume} {90}},\ \bibinfo {pages} {195428} (\bibinfo
  {year} {2014})}\BibitemShut {NoStop}%
\bibitem [{\citenamefont {Jin}\ \emph {et~al.}(2016)\citenamefont {Jin},
  \citenamefont {Li}, \citenamefont {Wang}, \citenamefont {Yu}, \citenamefont
  {Wan}, \citenamefont {Xu}, \citenamefont {Dai}, \citenamefont {Wei},\ and\
  \citenamefont {Guo}}]{JinJMCC}%
  \BibitemOpen
  \bibfield  {author} {\bibinfo {author} {\bibfnamefont {H.}~\bibnamefont
  {Jin}}, \bibinfo {author} {\bibfnamefont {J.}~\bibnamefont {Li}}, \bibinfo
  {author} {\bibfnamefont {B.}~\bibnamefont {Wang}}, \bibinfo {author}
  {\bibfnamefont {Y.}~\bibnamefont {Yu}}, \bibinfo {author} {\bibfnamefont
  {L.}~\bibnamefont {Wan}}, \bibinfo {author} {\bibfnamefont {F.}~\bibnamefont
  {Xu}}, \bibinfo {author} {\bibfnamefont {Y.}~\bibnamefont {Dai}}, \bibinfo
  {author} {\bibfnamefont {Y.}~\bibnamefont {Wei}}, \ and\ \bibinfo {author}
  {\bibfnamefont {H.}~\bibnamefont {Guo}},\ }\href {\doibase
  10.1039/C6TC04241D} {\bibfield  {journal} {\bibinfo  {journal} {J. Mater.
  Chem. C}\ }\textbf {\bibinfo {volume} {4}},\ \bibinfo {pages} {11253}
  (\bibinfo {year} {2016})}\BibitemShut {NoStop}%
\bibitem [{\citenamefont {Akimov}\ and\ \citenamefont
  {Prezhdo}(2013)}]{Akimov2013The}%
  \BibitemOpen
  \bibfield  {author} {\bibinfo {author} {\bibfnamefont {A.~V.}\ \bibnamefont
  {Akimov}}\ and\ \bibinfo {author} {\bibfnamefont {O.~V.}\ \bibnamefont
  {Prezhdo}},\ }\href@noop {} {\bibfield  {journal} {\bibinfo  {journal} {J.
  Chem. Theory Comput.}\ }\textbf {\bibinfo {volume} {9}},\ \bibinfo {pages}
  {4959} (\bibinfo {year} {2013})}\BibitemShut {NoStop}%
\bibitem [{\citenamefont {Akimov}\ and\ \citenamefont
  {Prezhdo}(2014)}]{Akimov2014Advanced}%
  \BibitemOpen
  \bibfield  {author} {\bibinfo {author} {\bibfnamefont {A.~V.}\ \bibnamefont
  {Akimov}}\ and\ \bibinfo {author} {\bibfnamefont {O.~V.}\ \bibnamefont
  {Prezhdo}},\ }\href@noop {} {\bibfield  {journal} {\bibinfo  {journal} {J.
  Chem. Theory Comput.}\ }\textbf {\bibinfo {volume} {10}},\ \bibinfo {pages}
  {789} (\bibinfo {year} {2014})}\BibitemShut {NoStop}%
\bibitem [{\citenamefont {Jaeger}\ \emph {et~al.}(2012)\citenamefont {Jaeger},
  \citenamefont {Fischer},\ and\ \citenamefont
  {Prezhdo}}]{Jaeger2012Decoherence}%
  \BibitemOpen
  \bibfield  {author} {\bibinfo {author} {\bibfnamefont {H.~M.}\ \bibnamefont
  {Jaeger}}, \bibinfo {author} {\bibfnamefont {S.}~\bibnamefont {Fischer}}, \
  and\ \bibinfo {author} {\bibfnamefont {O.~V.}\ \bibnamefont {Prezhdo}},\
  }\href@noop {} {\bibfield  {journal} {\bibinfo  {journal} {J. Chem. Phys.}\
  }\textbf {\bibinfo {volume} {137}},\ \bibinfo {pages} {22A545} (\bibinfo
  {year} {2012})}\BibitemShut {NoStop}%
\bibitem [{\citenamefont {Long}\ \emph {et~al.}(2017)\citenamefont {Long},
  \citenamefont {Prezhdo},\ and\ \citenamefont {Fang}}]{Long2017Nonadiabatic}%
  \BibitemOpen
  \bibfield  {author} {\bibinfo {author} {\bibfnamefont {R.}~\bibnamefont
  {Long}}, \bibinfo {author} {\bibfnamefont {O.~V.}\ \bibnamefont {Prezhdo}}, \
  and\ \bibinfo {author} {\bibfnamefont {W.}~\bibnamefont {Fang}},\ }\href@noop
  {} {\bibfield  {journal} {\bibinfo  {journal} {WIREs Comput. Mol. Sci.}\
  }\textbf {\bibinfo {volume} {7}},\ \bibinfo {pages} {e1305} (\bibinfo {year}
  {2017})}\BibitemShut {NoStop}%
\bibitem [{\citenamefont {Dvorak}\ \emph {et~al.}(2013)\citenamefont {Dvorak},
  \citenamefont {Wei},\ and\ \citenamefont {Wu}}]{dvorak2013origin}%
  \BibitemOpen
  \bibfield  {author} {\bibinfo {author} {\bibfnamefont {M.}~\bibnamefont
  {Dvorak}}, \bibinfo {author} {\bibfnamefont {S.-H.}\ \bibnamefont {Wei}}, \
  and\ \bibinfo {author} {\bibfnamefont {Z.}~\bibnamefont {Wu}},\ }\href@noop
  {} {\bibfield  {journal} {\bibinfo  {journal} {Phys. Rev. Lett.}\ }\textbf
  {\bibinfo {volume} {110}},\ \bibinfo {pages} {016402} (\bibinfo {year}
  {2013})}\BibitemShut {NoStop}%
\end{thebibliography}
\end{document}